\documentclass[preprint,12pt,authoryear]{elsarticle}

\usepackage{amssymb}
\usepackage{siunitx}
\usepackage{amsmath}
\usepackage[compact]{titlesec}
\usepackage{csquotes}

\titlespacing*{\subsection}
  {0pt}{\baselineskip}{\baselineskip}
\titlespacing*{\section}
  {0pt}{\baselineskip}{\baselineskip}
\titlespacing*{\subsubsection}
  {0pt}{\baselineskip}{\baselineskip}

\begin{document}

\begin{frontmatter}



\title{Synchrotron X-ray phase-contrast imaging of ultrasonic drop atomization}


\author[inst1]{Anunay~Prasanna\corref{cor1}}
\ead{aanunay@ethz.ch}
\author[inst1]{Luc~Biasiori-Poulanges}
\author[inst1]{Ya-Chi~Yu}
\author[inst4]{Hazem~El-Rabii}
\author[inst5]{Bratislav~Luki\'c}
\author[inst1]{Outi~Supponen}
\cortext[cor1]{Corresponding author: Institute of Fluid Dynamics, D-MAVT, Sonneggstrasse 3, ETH Z\"urich, 8092 Z\"urich, Tel: +41446326481}

\affiliation[inst1]{organization={Institute of Fluid Dynamics, D-MAVT},
            addressline={Sonneggstrasse 3, ETH Zurich}, 
            city={Zurich},
            postcode={8092}, 
            country={Switzerland}}

\affiliation[inst4]{organization={Institut Pprime, CNRS UPR 3346},
            city={Poitiers},
            postcode={86073}, 
            country={France}}  

\affiliation[inst5]{organization={European Synchrotron Radiation Facility},
            addressline={CS 40220}, 
            city={Grenoble},
            postcode={F-38043}, 
            country={France}}

\begin{abstract}
Ultrasonic atomization is employed to generate size-controllable droplets for a variety of applications. Here, we minimize the number of parameters dictating the process by studying the atomization of a single drop pending from an ultrasonic horn. Spatiotemporally resolved X-ray phase-contrast imaging measurements show that the number-median sizes of the ejected droplets can be predicted by the linear Navier-Stokes equations, signifying that the size distribution is controlled by the fluid properties and the driving frequency. Experiments with larger pendant water drops indicate that the fluid-structure interaction plays a pivotal role in determining the ejection onset of the pendant drop. The atomization of viscoelastic drops is dictated by extended ligament formation, entrainment of air, and ejection of drop-encapsulated bubbles. Existing scaling laws are used to explain the required higher input amplitudes for the complete atomization of viscoelastic drops as compared to inviscid drops. Finally, we elucidate the differences between capillary wave-based and cavitation-based atomization and show that inducing cavitation and strong bubble oscillations quickens the onset of daughter drop ejection but impedes their size control.
\end{abstract}




\end{frontmatter}


\section{Introduction}
\label{sec:intro}

Atomization is defined as the process of breaking up bulk liquid into smaller droplets. Ultrasonic transducers provide atomization at lower energy costs than mechanical atomizers, and therefore, ultrasonic atomization is typically the desired technique for most atomization applications. These include the preparation of specialty alloy powders \citep{lierke1967}, the creation of aerosols for pulmonal drug delivery \citep{taylor1997}, encapsulation in the food and pharmaceutical industry \citep{klaypradit2007}, emulsification processes \citep{Taha2020UltrasonicEmulsions} and chemical sonoreactors \citep{McCarogher2021}. Applications involving atomization require high throughput in order to create millions of droplets at a time \citep{Tsai2012}. Furthermore, the size distribution of the ejected droplets needs to be predictable and controllable to optimize the efficiency of the different applications \citep{Rajan2001, gogate2015}. 

Several attempts have been made to elucidate the physical mechanisms involved in ultrasonic atomization and to predict the resulting end products. \cite{Sollner1936} hypothesized that cavitation occurs in a liquid film subjected to ultrasound excitation, and that hydraulic shock generation in the film leads to the ejection of daughter drops. On the other hand, \cite{Lang1962} showed that capillary waves are first formed on the surface of the liquid film, and the associated hydrodynamic instabilities are the primary mechanisms for the ejection of daughter drops. In that case, the number-median diameter of the ejected drops, $d_e$, could be predicted as a constant fraction of the capillary wavelength \citep{Lang1962}. More recent studies have increased the parameter space and produced empirical correlations to predict the mean droplet size in ultrasonic atomization \citep{Rajan2001, Ramisetty2013}. The currently accepted theory is the conjunction theory or the combined cavitation-capillary theory, which states that cavitation events contribute to the creation of capillary waves on the surface, which then rupture to form daughter droplets. A multitude of studies have demonstrated the conjunction theory in practice \citep{Tomita2014, Simon2015, Cailly2023}.

Interestingly, a number of state-of-the-art applications are attempting to employ high-frequency excitation (in the \si{\mega\hertz} range) on drops attached to acoustic transducer surfaces to produce daughter droplets \citep{Tsai2012, Simon2015}. The acoustic excitation of pendant drops has mainly been studied from a fundamental point of view. Research has been dedicated to estimating typical oscillation behavior and frequencies \citep{Strani1984, Bostwick2014, Chang2015}, surface capillary wave formation \citep{Wilkes1997, Vukasinovic2007, Tan2010}, and the hysteretic response of the drops to the applied forcing \citep{DePaoli1995, Wilkes1999b}.  One of the first notable attempts to explain drop atomization behavior was made by \cite{Goodridge1997}, having performed experiments on millimetric sessile drops at low-frequency excitations ($f_d = 20 - 60\;\si{\hertz}$). Their drops showed nonlinear wave-amplitude behavior, with the threshold amplitude for the ejection of daughter droplets being a function of the applied frequency and the surface tension or the viscosity, depending on whether the drops were inviscid or viscous respectively. \cite{James2003} studied the atomization characteristics of a sessile drop at a higher driving frequency ($f_d = 1000 \;\si{\hertz}$). They found that the atomization depended on the coupled fluid-structure interaction of the drop and the vibrating surface, with the drop bursting completely into daughter droplets, provided that the natural frequency of the combined structure and drop was in resonance with the driving frequency. Further studies have investigated the ejection mechanism \citep{James2003_2, Vukasinovic2007, Deepu2013} and the required threshold amplitudes for ejection \citep{Deepu2018}. 

However, most of these studies are typically undertaken at low frequencies (usually a multiple of the natural frequency of the drop) and do not comprehensively report on higher frequency daughter drop ejection characteristics, making them difficult to compare with actual applications. Furthermore, based on the method of application of ultrasonic excitation, there seem to exist disagreements on what is the dominant mechanism of droplet ejection. Previous experiments indicate that the effects are case-specific, with the ejected drop size distribution and the throughput being dictated by a large parameter space, including fluid properties and the characteristics of the applied ultrasound. \cite{antonevich1959} hypothesized from experiments that the stochastic nature of cavitation would lead to a wider range of ejected drop sizes as compared to having ejection from the rupture of capillary waves only. Recent studies have further attempted to quantify the specific role of cavitation in liquid film atomization \citep{Zhang2021, Zhang2022}. However, there still remains some debate on the distinct contributions of capillary waves and cavitation-related activities with respect to the ejection process, and therefore, there is a need to outline and compare the roles of both mechanisms. The complexity involved in resolving all the spatiotemporal scales of the problem implies that the current studies used for predicting ejected drop sizes are only reduced-order solutions. We believe that with simple configurations, this study can shed light not only on which is the dominant mechanism for drop atomization, but also on the end products of the atomization process based on the atomization mechanism. This should also enable us to define which mechanism is better suited for atomization applications.


In this work, the phenomenology of drop atomization is underlined by utilizing advanced experimental techniques. A pendant drop is attached to the tip of an ultrasonic horn capable of nucleating vapor bubbles within a liquid volume \citep{luc2023}. By using different configurations, we initiate capillary waves on the drop interface, both with and without cavitation. To further simplify the problem, we reduce the number of parameters to analyze by fixing the input amplitude and driving frequency of the ultrasonic horn. Time-resolved X-ray phase-contrast imaging coupled with conventional shadowgraphy is employed to elucidate the underlying mechanisms involved. This allows us to overcome the overlapping problems of dense droplet sprays and provides a detailed visualization of surface instabilities, cavitation, and ejection mechanisms underlying the drop fragmentation process.

The outline of the paper is as follows. First, we describe the experimental methods. We then provide a qualitative description of the ejection process for water droplets and quantify the ejected drop sizes. In addition, 2\% chitosan, which is a viscoelastic fluid, is then used to make pendant droplets that are subjected to the same periodic forcing, and a possible explanation for their ejection dynamics is provided. Chitosan was chosen as our fluid with viscoelastic properties due to its relevance in ultrasound-enhanced bioadhesive and transdermal drug delivery \citep{ma2022} which are topics we work with extensively. Further configurations of droplets are tested to study the effects of cavitation on vibration-induced pendant drop atomization in greater detail. 

\section{Experimental Apparatus and Methodology}
\label{sec:meth}

\begin{figure*}[t]
    \centering
    \includegraphics[width=\linewidth]{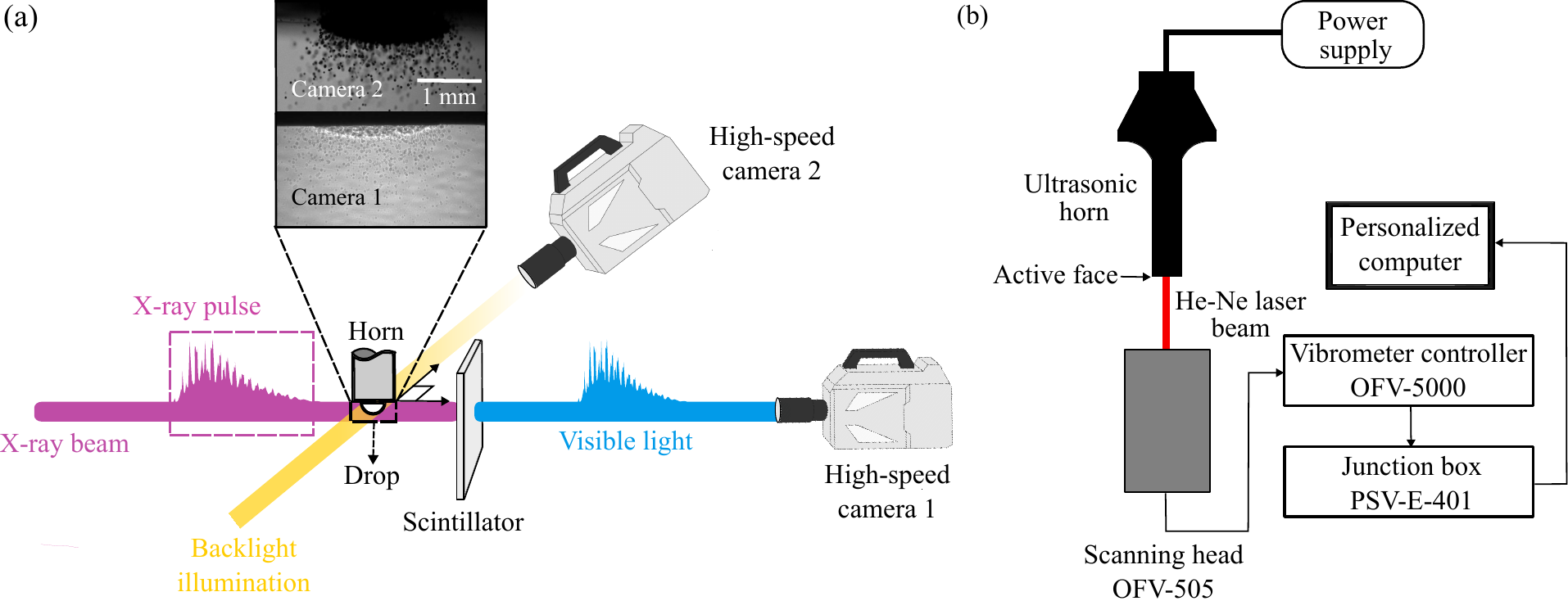}
    \caption{(a) Experimental setup depicting the high-speed imaging approaches employed: Synchrotron X-ray phase-contrast imaging (Camera 1) and shadowgraphy (Camera 2) (b) Laser scanning vibrometry setup to determine the response of the active face of the ultrasonic horn}
    \label{fig: exp_setup}
\end{figure*}

\subsection{Pendant drop and ultrasonic excitation}

Drops of water (at volumes of $V = 50\;\text{and}\; 100\;\si{\micro\liter}$) and 2\% chitosan in a hydrochloric acid (HCl) and water solution are used for the experiments. The preparation of the 2\% chitosan solution is outlined in \ref{app:chitosan}. The drops are then placed on the active face of the horn with the help of a microsyringe. Due to the expected high strain rates, we select a viscosity value closer to the infinite shear rate value expected for the 2\% chitosan ($\mu_\infty = 100 \;\si{\milli\pascal\second}$), as estimated by shear rheometry \citep{Cho2006}. Modeling the nonlinear, viscoelastic nature of chitosan in this case was non-trivial and beyond the scope of the paper.


A 1/2" diameter ultrasonic horn (Branson Ultrasonic Sonifier SFX550, 550~\si{\watt}) with a driving frequency of 20~\si{\kilo\hertz} is operated at 40\% of maximum amplitude, which corresponds to a maximum peak-to-peak displacement of $A_\mathrm{pp} = 57 \;\si{\micro\meter}$. The horn adjusts its power output based on the viscous dissipation of the liquid to provide the same peak-to-peak displacement for a given amplitude percentage irrespective of the liquid attached to its surface. The horn has a transient period of approximately 16~\si{\milli\second} before the maximum amplitude has fully developed and the acoustic field has set in.  

\subsection{High-speed imaging}

The experiments are carried out at the ID19 beamline of the European Synchrotron Radiation Facility \citep{weitkamp2010}.  The polychromatic X-ray beam generated by a long-period undulator set to a 20-\si{\milli\meter} gap is used to probe the fast ultrasound-induced atomization dynamics. The (partial) spatial coherence properties and the high flux of the X-ray beam are leveraged to improve the contrast between dissimilar phases and resolve fluid interfaces while preserving the overall geometrical representation in direct space \citep{luc2023}. The X-ray spectrum used to illuminate the sample is filtered with mandatory optical elements along the 145-\si{\meter}-long vacuum flight tube to provide heat-load moderation (2.2-\si{\milli\meter} thick diamond window and a series of thin carbon and beryllium windows). The X-ray detector consists of a 1-\si{\milli\meter} thick LuAg:Ce (Ce-doped Lu$_{3}$Al$_{5}$O$_{12}$) scintillator optically coupled to the Photron SA-Z ultra-fast camera (High-speed camera 1 in Fig.~\ref{fig: exp_setup}(a)) equipped with $2.1 \times$ magnification (100:210 Hasselblad tandem optic) and provided a  pixel size of 9.52~\si{\micro\meter}. Due to a micrometric source size and (almost) parallel illumination, the penumbral blur is orders of magnitude below the pixel size. The detector assembly is positioned 5.5~\si{\meter} downstream of the sample, ensuring that the propagation-based interference between transmitted X-rays results in an increased edge contrast due to (partial) spatial beam coherence while fulfilling the near-field condition \citep{Wilkins1996}. 

Shadowgraphy is performed simultaneously on an axis perpendicular to the X-ray phase-contrast imaging, which is captured by a Photron FASTCAM NOVA S12 (High-speed camera 2 in Fig.~\ref{fig: exp_setup}(a)). Both cameras were set at recording frame rates of 80~\si{\kilo\hertz}. The inset of Fig.~\ref{fig: exp_setup}(a) clearly depicts the differences between the X-ray and shadowgraph images. Every ejected daughter droplet is in focus in the phase-contrast images contrary to the shadowgraphs. The surface and the interior of the droplet are clearly visible in the phase-contrast images, which allows us to evaluate qualitative differences between the test cases in detail. The clarity of the phase-contrast images enables us to evaluate ejected droplet size distributions accurately - an estimate that would at best be approximate with the shadowgraphs. Further differences between the images from the two techniques can be evaluated by the reader from the Supplementary Videos provided for a water droplet with contact radius, $R_c = 1.4\;\si{\milli\meter}$.  

\subsection{Laser scanning vibrometry}

To assess the dynamical behavior of the active face of the ultrasonic horn to which the pendant drop is attached, laser scanning vibrometry is employed. It is an optical method that uses the Doppler effect to measure the velocity of the surface \citep{drain1980}. The experimental setup is depicted in Fig.~\ref{fig: exp_setup}(b). A Polytec Scanning Vibrometer (PSV-400) with a controller (OFV-5000), capable of measuring frequencies up to 1 \si{\mega\hertz}, is used for the measurement. The scanning head (OFV-505) has a Helium-Neon (He-Ne) laser with a  wavelength of $\lambda = 633$~\si{\nano\meter}. The VD-09 decoder with a maximum range of 1000 \si{\milli\meter\per\second\per\volt} was used, since the typical velocity range expected for the horn was around 0 - 10 \si{\meter\per\second}. Grid points to be scanned by the vibrometer are defined on the active face of the horn. The ultrasonic horn is activated at different horn amplitudes ($A = 30\%,\; 40\% \;\mathrm{and}\; 50\%$), and after the transient period of the horn, the vibrometer is set to scan the defined grid points. This is used to obtain the frequency response and the velocity amplitude of the active face of the horn, with the velocity amplitude at the different scan points dictating the dominant mode shape of the active face of the horn.

\section{Results}
\label{sec:res}

\subsection{Qualitative description of ejection process for water drops}
\label{sec:water_drop}

\begin{figure*}[t]
    \centering
    \includegraphics[width=\linewidth]{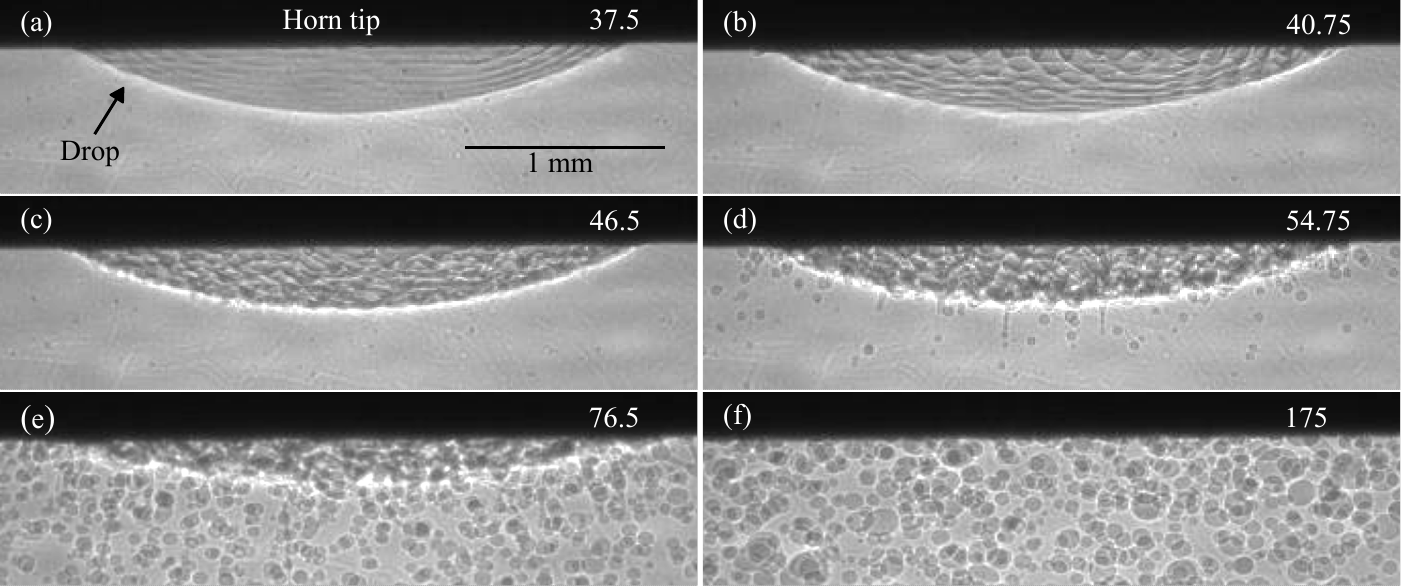}
    \caption{The different mechanisms experienced by a pendant water drop with $R_c = 1.4 \; \si{\milli\meter}$ subjected to periodic forcing with a driving frequency of $f_d = 20 \; \si{\kilo\hertz}$. The snapshots of X-ray phase-contrast images are labeled with their non-dimensional times, $t/T_d$, where $T_d = 1/f_d$ is the time period of the driving ultrasonic excitation. The field of view is cropped for improved visualization of the pendant drop. (a) Harmonic, axisymmetric waves; (b) Subharmonic, azimuthal waves; (c) A "lattice" mode, formed by the interaction of different wave modes (d) Ejection onset of daughter droplets and ligaments along the drop interface; (e) Chaotic ejection of multiple daughter droplets and ligaments; (f) Complete atomization of the pendant drop with only daughter droplets in the field of view.}
    \label{fig: wave_modes}
\end{figure*}

\begin{figure*}[t]
    \centering
    \includegraphics[width=\linewidth]{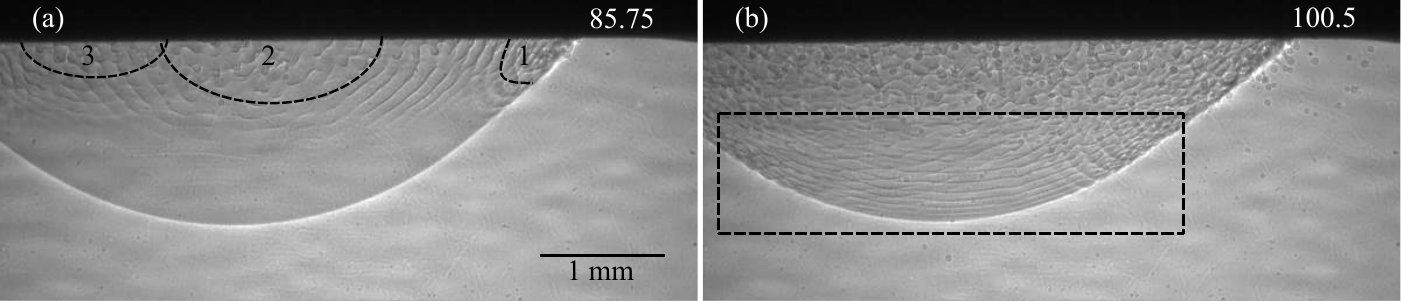}
    \caption{Wave patterns on a pendant water drop with $R_c = 2.8 \; \si{\milli\meter}$. The snapshots of X-ray phase-contrast images are labeled with their non-dimensional times, $t/T_d$. (a) Subharmonic semi-circular waves originate from \enquote{point sources}, where the sources are depicted by numbers, and the waves are marked by black dotted lines; (b) Both subharmonic semi-circular waves and harmonic waves are visible as indicated by the black dotted box. Droplet ejection is seen from the location of the \enquote{point sources} near the contact line.}
    \label{fig: point_source}
\end{figure*}

Vibration-based atomization of water drops of volumes $V = 50 \;\text{and}\; 100 \; \si{\micro\liter}$, corresponding to contact radii $R_c = 1.4\;\si{\milli\meter}$ and $R_c = 2.8\;\si{\milli\meter}$ respectively, is investigated. From here on, the contact radius will be used to distinguish the drops, as this can be carried over to other drop configurations used later. Exciting the drop under periodic forcing leads to the formation of capillary waves as reported by other studies for both pendant and sessile drops \citep{Wilkes1997, Goodridge1997, Biasiori-Poulanges2022Video:Fragmentation}. The frequency content of wave packets can be measured in a manner similar to \cite{Vukasinovic2007_JFM}, by taking the intensity variation of a pixel on the image and calculating its power spectral density over time. The results have been briefly summarized here, with a detailed description available in \cite{James2003} and \cite{Vukasinovic2007_JFM} for sessile water drops.

\begin{figure*}[t]
    \centering
    \includegraphics[width=\linewidth]{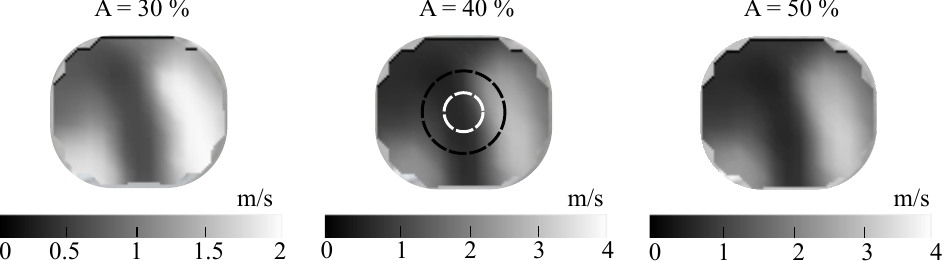}
    \caption{The velocity amplitude of the dominant mode of the active face of the ultrasonic horn at three different amplitude percentages obtained by laser scanning vibrometry. The two dotted circles drawn on the shape of $A=40\%$ show the expected positions of the pendant drops with $R_c = 1.4\;\si{\milli\meter}$ (white) and $R_c = 2.8\;\si{\milli\meter}$ (black). It can be noticed that the high amplitude regions on the horn face behave as \enquote{point sources} for the pendant drop with $R_c = 2.8\;\si{\milli\meter}$}
    \label{fig: point_source_horn}
\end{figure*}

The atomization of the drop can be characterized into six stages as visualized in Fig.~\ref{fig: wave_modes}. First, the drop radially oscillates for a while, before axisymmetric standing waves become visible in Fig.~\ref{fig: wave_modes}(a). These waves require minimal excitation amplitude to exist \citep{James2003} and are harmonic in nature, with the frequency content of the wave packets being the same as the driving frequency ($f_d = 20\;\si{\kilo\hertz}$). As the excitation amplitude increases, azimuthal waves are developed at the contact line of the drop along the horn (seen in Fig.~\ref{fig: wave_modes}(b)). These waves are subharmonic and have a frequency content equivalent to half the driving frequency, implying that they correspond to the classical Faraday instability \citep{faraday1831}. They move slowly downwards along the surface, from the contact line to the tip of the drop, and interact with the existing harmonic, axisymmetric waves to create higher-order spatial modes on the drop. As the amplitude of the waves on the drop increases, crests and troughs on the surface grow progressively and become more visible. Over time, the interaction of different wave modes leads to the creation of a time-dependent "lattice" mode on the surface of the drop as seen in Fig.~\ref{fig: wave_modes}(c). The process is chaotic, and it is difficult to distinguish a single dominant frequency for this phase of drop vibration. Fig.~\ref{fig: wave_modes}(d) shows the onset of ejection on the primary drop surface, leading to the formation of several daughter droplets. Here, we define the onset of ejection as when there are ejection sites over the entire pendant drop interface. Ejection of daughter droplets occurs due to the collapse of troughs on the pendant drop interface. Sometimes ligaments are ejected instead of droplets, which are then susceptible to the Rayleigh-Plateau instability, leading to the formation of satellite droplets \citep{Eggers2008}. Over time, the number of daughter droplets and ligaments ejected increases as seen in Fig.~\ref{fig: wave_modes}(e), before the pendant drop completely atomizes (see Fig.~\ref{fig: wave_modes}(f)) and only the ejected daughter droplets exist in the field of view. 

The overall atomization dynamics of the pendant drop are similar when the volume is increased ($R_c = 2.8\;\si{\milli\meter}$) as seen in Fig.~\ref{fig: point_source}. Axisymmetric harmonic waves are created first in the droplet. However, unlike in the case of $R_c = 1.4\;\si{\milli\meter}$, some locations on the drop have already generated subharmonic waves before the axisymmetric waves have fully set in. Three such locations are depicted in Fig.~\ref{fig: point_source}(a). These locations behave as \enquote{point sources}, generating subharmonic waves that continue to interact with the harmonic waves, clearly influencing the ejection process as shown in Fig.~\ref{fig: point_source}(b). The first daughter droplets are ejected from the locations of the point sources, as opposed to the whole drop surface as for the case of $R_c = 1.4\;\si{\milli\meter}$.

Since the point sources are not created by the fluid itself, it can be conjectured that the ultrasonic horn had a role to play in the creation of the point sources. Investigations of the frequency response and the mode shapes of the horn face were carried out using laser scanning vibrometry. Fig.~\ref{fig: point_source_horn} depicts the velocity amplitudes of the active face of the horn for three different horn amplitude percentages. The contour represents the dominant mode of the horn at its fundamental frequency and indicates that there exist regions of high localized displacement with nearly twice the amplitude of the lower displacement regions on the horn face. While the amplitudes of all the regions in the horn are sufficient to create both harmonic and subharmonic waves \citep{Kumar1994}, it is evident from our experiments that the subharmonic waves need a longer time to set in than the harmonic waves. With localized regions of high amplitude on the active face of the horn, it is easier to cross the threshold required to generate subharmonic waves on the drop surface. Therefore, these locations on the horn face seem to behave as "point sources", allowing for a quicker transition between the different stages of drop ejection (harmonic waves - subharmonic waves - "lattice" mode - ejection onset). This discussion clearly indicates that the fluid-structure interaction plays an important role in the atomization process, particularly in determining the onset of ejection, and the regions on the primary drop from where the ejection of daughter droplets begins.


\subsection{Ejected droplet size distribution}

The daughter droplet sizes ejected by the pendant water drop can be counted from the frames of the phase-contrast images. A Canny edge detection algorithm was employed to estimate the droplet size distribution from selected frames of the different test cases \citep{Canny1986}. A few shadowgraph frames at the same instants as the phase-contrast images were also processed to estimate if the size distributions obtained were consistent. Due to the necessity to exclude the primary pendant drop from the edge detection algorithm so as to not detect the change in image contrast generated by the surface waves, the pendant drop was masked and excluded from the field of view. For the case of $R_c = 1.4 \; \si{\milli\meter}$, 80\% of the captured field of view can be isolated to count the droplets, while for $R_c = 2.8\;\si{\milli\meter}$, 60\% of the field of view is used. 

In the earlier stages of ejection where fewer droplets exist in a frame, nearly 90\% of the ejected droplets are counted correctly. The miscounted droplets are usually the ones that are still near the surface of the primary drop, and so we expect to count them anyway at a later instant. As the fraction of droplets in a frame increases, the performance of the edge detection algorithm decreases. As soon as less than 50\% of the droplets in a frame are counted correctly, we stop our evaluation. With this criteria, we evaluate 620 frames for the case of $R_c = 1.4\;\si{\milli\meter}$ and 220 frames for $R_c = 2.8\;\si{\milli\meter}$. This lets us count millions of droplets for both cases and we believe that is sufficient to correctly estimate the ejected drop size distribution. The error associated with the edge detection algorithm itself is approximated by evaluating the intersection between a defined ground truth and the estimate from the algorithm \citep{Lopez2013}. Doing so for several cases and averaging the associated error, allows us to estimate the overall error as 1 px, which corresponds to 9.52 \si{\micro\meter}. 

\begin{figure*}[p]
    \centering
    \includegraphics[width=0.8\linewidth]{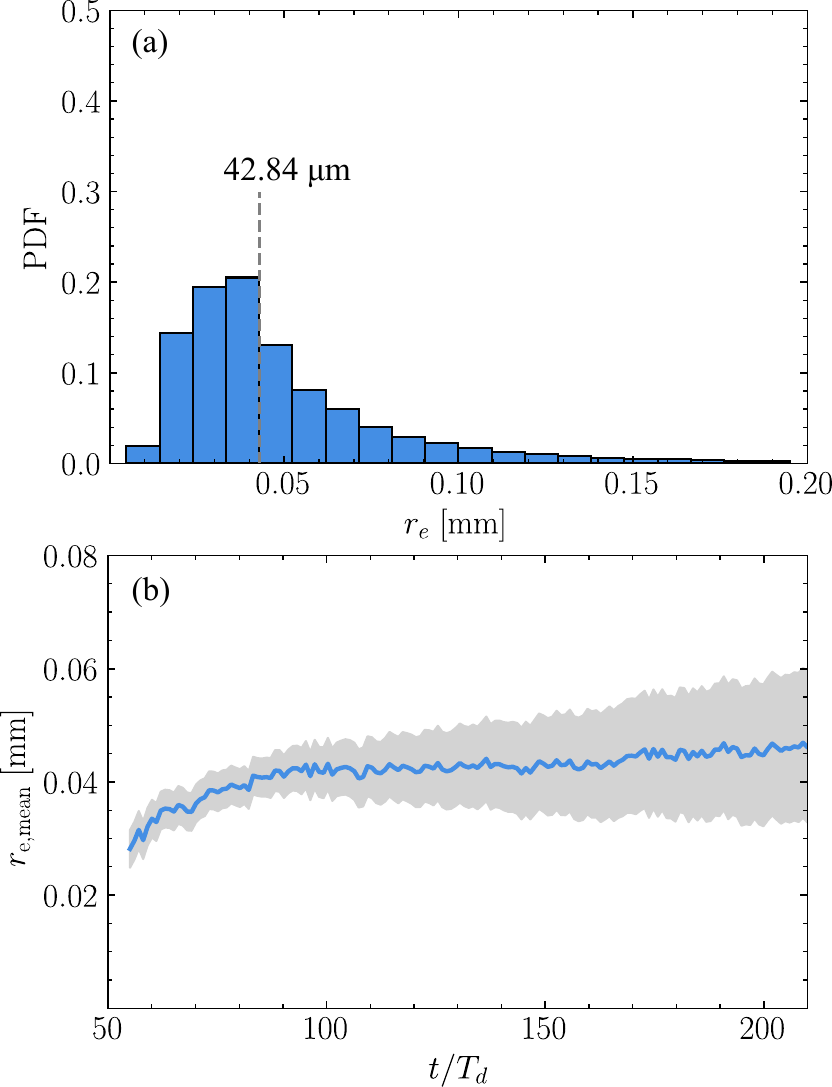}
    \caption{(a) The probability distribution of the radius of the ejected droplets for a pendant water drop with $R_c = 1.4 \; \si{\milli\metre}$ plotted for the driving cycles, $55 \leq t/T_d \leq 210$. The mean (indicated by the dashed line) and median radius of the ejected droplets are 42.84 $\pm$ 9.52 \si{\micro\meter} and 33.1 $\pm$ 9.52 \si{\micro\meter} respectively, where the standard deviation is the error obtained from edge detection. (b) The evolution of the mean radius of the ejected droplets for $R_c = 1.4\;\si{\milli\meter}$ evaluated per frame, plotted for $55 \leq t/T_d \leq 210$. The shaded region shows the error associated with edge detection, with the error increasing over time due to the decreasing percentage of ejected droplets counted.}
    \label{fig: drop_size_1}
\end{figure*}

The droplet size distribution for $R_c = 1.4\;\si{\milli\meter}$ is depicted in Fig.~\ref{fig: drop_size_1}(a). The ejected droplets are polydisperse and can be modeled as a Gaussian distribution. \cite{Lang1962} predicted that the median drop size scales as $d_e = 0.34\lambda_f$, where $\lambda_f$ corresponds to the expected Faraday wavelength given by
\begin{align}
    \lambda_f = \left(\frac{8\pi\sigma}{\rho f_d^2}\right)^{1/3}
    \label{eq:lambda}
\end{align}
where $\sigma$ is the surface tension between the liquid and air, $\rho$ is the density of the fluid, and $f_d$ is the driving frequency. For $f_d = 20\;\si{\kilo\hertz}$, this corresponds to a value of $r_{e, \mathrm{th}} = 28.11\;\si{\micro\meter}$, with our experimental median being $r_{e, \mathrm{exp}} = 33.1\;\si{\micro\meter}$, which is well within the error associated with edge detection, and shows a good correspondence with Eq.(~\ref{eq:lambda}).   

Fig.~\ref{fig: drop_size_1}(b) plots the evolution of the mean radius of the ejected droplets evaluated per image frame. Isolating the distribution over different time periods of forcing, it can be seen that the mean droplet size of the distribution initially increases over time. Smaller droplets closer to the median size are ejected during the initial cycles, with larger droplets being ejected over later cycles. To estimate the ejected mean droplet size theoretically, one would have to evaluate the eigenfrequencies of the capillary waves of a partially wetting drop subject to (pointlike) forced oscillations. However, modeling this is nontrivial and beyond the scope of this study.  

It should be noted that Eq.~(\ref{eq:lambda}) is accurate in predicting the ejected droplet sizes for low amplitude waves and where the viscosity of the atomized fluid does not play an important role \citep{Rajan2001}. However, the polydisperse distribution and the chaotic spike-like structures on the surface of the primary drop indicate that nonlinear effects, similar to the ones seen for lower oscillation frequencies \citep{Goodridge1997}, are significant, especially at later stages of ejection. Therefore, while predicting the ejected droplet sizes theoretically is challenging, it is still interesting to note that Eq.~(\ref{eq:lambda}) provides good estimates of the ejected droplet size.

\subsection{Ejection behavior of a drop with 2\% chitosan solution}
\label{sec:chitosan}

\begin{figure*}[p]
    \centering
    \includegraphics[width=0.85\textwidth]{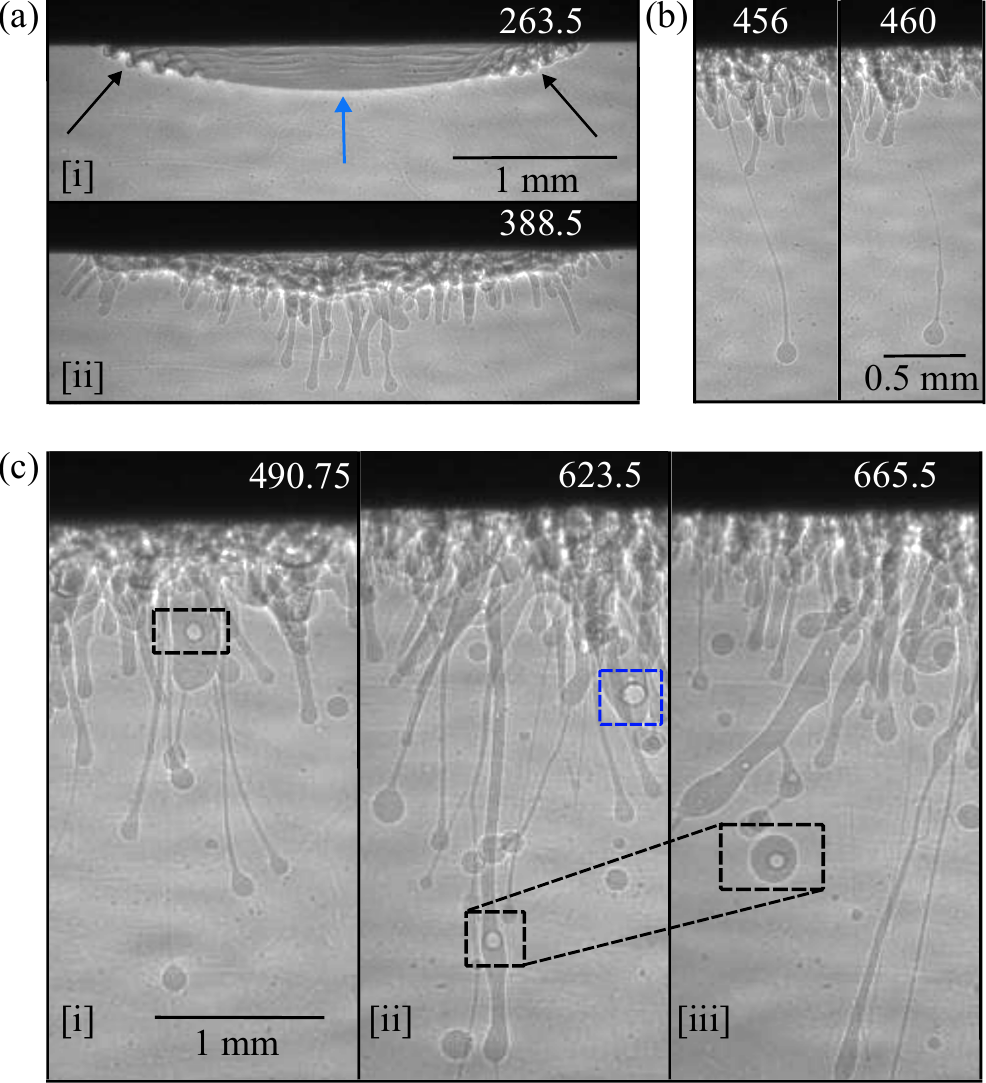}
    \caption{The dynamics of a pendant 2\% chitosan drop with $R_c = 1.6 \; \si{\milli\meter}$ subjected to a periodic vibration of $f_d = 20 \; \si{\kilo\hertz}$. The snapshots of the X-ray phase-contrast images are labeled with their non-dimensional times, $t/T_d$. (a) Evolution of the different wave modes on the drop: [i] shows both harmonic (blue arrow) and subharmonic waves (black arrows). [ii] shows the formation of long, viscoelastic ligaments; (b) A single ligament and its subsequent breakup from the base; (c) Different instances of air entrainment within the primary drop and its ligaments: [i] intertwining ligaments entrap an air bubble depicted within the black box [ii] Interacting ligaments entrain more air within the primary drop (dark blue box), while large amplitude disturbances can break the surface of ligaments to entrap air in them (black box). [iii] The air entrained in ligaments can be ejected along with daughter droplets. Here, the entrained air in the ligament shown in [ii] is ejected as a drop-encapsulated bubble (black box).}
    \label{fig: chitosan_waves_ejection_entrainment}
\end{figure*}

As mentioned previously, the 2\% chitosan solution is highly shear-thinning and shows considerable viscoelastic behavior. Due to the high strain rates involved with the majority of the flow ($\dot{\gamma} \sim 2 \pi f_d$), an effective viscosity closer to the infinite shear rate viscosity will be employed for further analysis \citep{Evans2007}. 

The dynamics are elucidated for a drop with $R_c = 1.6 \;\si{\milli\meter}$, corresponding to a volume of 50 \si{\micro\liter}. The first noticeable difference is that the drop initially spreads on the active face of the horn, indicating that these configurations have dynamic contact lines \citep{DePaoli1995}. This brings the drop in contact with the localized regions of high displacement on the face of the horn, which leads to the existence of both harmonic and subharmonic waves on the drop surface as shown in Fig.~\ref{fig: chitosan_waves_ejection_entrainment}(a). The subharmonic waves arise from the sides of the drop interface where, as shown previously in Fig.~\ref{fig: point_source_horn}, higher amplitudes of the horn are expected. Here, the drop is shown at $t/T_d = 263.5$, which is a much later driving cycle than for the water drop with $R_c = 1.4\;\si{\milli\meter}$, where the whole process of surface wave formation to the onset of ejection takes place between $35 \leq t/T_d \leq 55$. For parametric instabilities, it has been shown theoretically that a viscous fluid stabilizes the interface, and it takes a much higher amplitude to trigger the surface instabilities as compared to an inviscid fluid \citep{Kumar1994, Ebo2016}. Given the highly viscous nature of the 2\% chitosan solution, it is unsurprising that it takes much longer for the different wave modes to set in than for water. 

The surface waves form craters and spikes similar to the water drop. However, no daughter droplets are ejected, which is elucidated further in Section~\ref{sec: thresh_acc}. Indeed, the collapse of troughs leads to the formation of ligaments that oscillate with the horn face as depicted in Fig.~\ref{fig: chitosan_waves_ejection_entrainment}(a). The pendant drop shows a highly viscoelastic behavior, which can be qualitatively characterized by the Deborah number, $\mathrm{De} = \kappa/\tau_f \sim \kappa/T_d$, where $\kappa$ is the relaxation time for the 2\% chitosan solution. For the present case, $\mathrm{De} > 1$ at all times as the relaxation time of the 2\% chitosan \citep{Cho2006} is much higher than the flow time scale, which indicates that the elastic effects of chitosan are relatively important \citep{bird1987}. In the initial cycles, it is conjectured that the strain rates within the ligaments are quite low, thus keeping the stress in the ligaments below the yield stress and allowing them to react to the oscillation of the ultrasonic horn. Due to the high degree of elasticity, the axial stresses and the strain rates within the ligaments increase, unbounded in time, making these ligaments extend and elongate for a large number of driving cycles \citep{Mckinley}. The jetting of the ligaments or drops is an impulsive process, implying that the collapse of troughs needs to be powerful enough to cross a  minimum velocity threshold for ejection to occur \citep{Vukasinovic2007}. Once this threshold has been exceeded, the long ligaments break up from the base rather than from the tip (see Fig.~\ref{fig: chitosan_waves_ejection_entrainment}(b)), as has also been reported for liquids of higher viscosity \citep{Goodridge1997, Vukasinovic2007}. Due to an elasto-inertial-capillary balance on individual ligaments, certain ligaments can also experience breakup from the middle when the axial stress stabilizing the ligament is lower than the force exerted by surface tension \citep{Chang1999}.  Furthermore, the high resistive stresses in these ligaments persist even after breakup making the ligaments quite stable to the Rayleigh-Plateau instability \citep{driessen2013}. Characteristic viscous and viscoelastic effects such as "gobbling" drops \citep{Clasen2009} and "beads-on-a-string" \citep{Ardekani2010} are also visible in some of the ligaments as seen in Fig.~\ref{fig: chitosan_waves_ejection_entrainment}.

Over time, the primary drop starts entrapping ambient air from the surroundings as depicted in Fig.~\ref{fig: chitosan_waves_ejection_entrainment}(c). The entrapped bubble(s) can oscillate with the primary drop for several cycles, and later break up to form multiple bubbles and enhance mixing within the primary drop. Three mechanisms leading to entrainment have been identified and are described as follows.

Due to the viscoelasticity of the ligament, some of the ejected daughter droplets recoil, provided they are ejected during the negative phase of the ultrasonic horn excitation. Since the ejected droplets have a radius of $\mathcal{O}(10^{-4})\;\si{\meter}$, the effect of gravity is negligible. Therefore, the existing inertia in the ejected droplets pulls them back towards the primary vibrating drop. These droplets impact the primary drop and entrain air, similar to larger drops impacting liquid pools as described by other studies \citep{Oguz1990, Oguz1995}. 

Entrainment also occurs when there are multiple ligaments close by and the negative phase of the horn causes these ligaments to swirl and entwine each other (this is the case for Fig.~\ref{fig: chitosan_waves_ejection_entrainment}(c)[i]). If the time scale of rupture of the air film is larger than the time scale of coalescence between the two ligaments, air can be entrapped. A similar mechanism has been noticed for studies involving entrainment in vibrating liquid-filled vessels \citep{Obenauf2022} and in piezoelectric inkjet printing \citep{deJong2006AirPrintheads}.

Large amplitude disturbances on the extended ligaments break the surface of the ligament, which can then curl up and entrain air within the ligament itself in a mechanism similar to other breaking surface waves \citep{Kiger2011}. This is depicted in Fig.~\ref{fig: chitosan_waves_ejection_entrainment}(c)[ii]. These ligaments then jet the entrained air along with a droplet to form drop-encapsulated bubbles that can remain stable for quite a long time, even up to several \si{\milli\second} in some cases (see Fig.~\ref{fig: chitosan_waves_ejection_entrainment}(c)[iii]), before the entrapped air coalesces with the surrounding, leaving just the ejected droplet. A detailed discussion on the benefits and adverse effects of having entrainment in such a configuration is provided in Section~\ref{sec:disc}.

\subsection{Threshold acceleration for droplet ejection}
\label{sec: thresh_acc}

The applied acceleration at which the onset of ejection takes place is defined as the threshold acceleration of ejection. The experiments show that the chitosan drop does not initially eject daughter droplets, unlike water drops, and is only capable of ejecting ligaments. \cite{Goodridge1997} and \cite{Vukasinovic2007} have also provided similar experimental results for glycerin-water mixtures, which have higher viscosities than pure water. Here, we try to explain the reason for this behavior using the explanation provided by \cite{Goodridge1997}, which is briefly summarized below.

For inviscid fluids, ejection occurs when the height of the surface waves roughly equals their wavelength, $h \sim \lambda$. Considering that the only counteracting effect against the input acceleration is the surface tension, $\sigma$, the critical displacement amplitude (assuming that $h_{\mathrm{cr}} \sim a_{\mathrm{th}}/\omega^2$) required to eject droplets can be given as
\begin{equation}
    h_{\mathrm{cr}} \sim \left(\frac{\sigma}{\rho}\right)^{1/3}\omega^{-2/3}
    \label{eq:disp_amp_inviscid}
\end{equation}
Using the material properties of water and substituting $\omega = 2\pi f_d$ gives the critical displacement for water as $h_\mathrm{cr} \sim 4 \;\si{\micro\meter}$, which is reached by our horn with $A_\mathrm{max} = 28.5\;\si{\micro\meter}$ for 40\% amplitude. Therefore, pendant water drops are able to eject daughter droplets quite easily in this configuration. For viscous fluids, the power applied to the system is balanced by the viscous dissipation of the fluid. Performing a similar scaling analysis as before, but now using viscosity instead of the surface tension gives 
\begin{equation}
    h_{\mathrm{cr}} \sim \left(\frac{\mu}{\rho}\right)^{1/2}\omega^{-1/2} 
    \label{eq:disp_amp_viscous}
\end{equation}
As stated before, using an effective viscosity for the 2\% chitosan solution drop ($\mu_\mathrm{eff} \approx 100 \;\si{\milli\pascal\second}$) and setting $\omega = 2\pi f_d$, gives the critical displacement value as $h_\mathrm{cr} \sim 25 \; \si{\micro\meter}$, which is very close to the maximum amplitude of our ultrasonic horn. This amplitude cannot be achieved in the transient period of the horn and therefore, any ejection processes from the drop can only take place after the transient period has passed and the acoustic field has been fully developed in the pendant drop. Therefore, the ejection process here sets in much later and is probably dictated by other events as well, such as the increase in the surface area occupied by the drop under the horn, and the entrainment of air within the primary drop that can further enhance the ejection of ligaments.   



\subsection{Effects of cavitation on drop ejection}
\label{sec:cav_ejection}

Here, we would like to compare cavitation-induced ejection with respect to only having capillary wave ejection by generating cavitation or cavitation-related activities inside the drop. Time-resolved X-ray phase-contrast imaging proved extremely beneficial in deducing cavitation inception inside drops and enabled us to evaluate the intrinsic bubble dynamics inside the liquid volume. We tested two different configurations to answer the changes induced by cavitation to the onset of ejection and the ejected droplet sizes. We confined water drops between the ultrasonic horn and a flat surface. The rigid confining surface at the bottom, along with the large impedance mismatch at the lateral gas-liquid interface, acts as a strong reflector for the incoming waves and can lead to cavitation within the trapped drop \citep{Moussatov2005, Fatjo2011}. Next, we trapped bubbles inside a drop to study specifically how the bubble oscillations and accelerations contribute to the ejection process.

\subsubsection{Confined water drops}
\label{sec:confined_water_drop}

\begin{figure*}[p]
    \centering
    \includegraphics[width=0.75\textwidth]{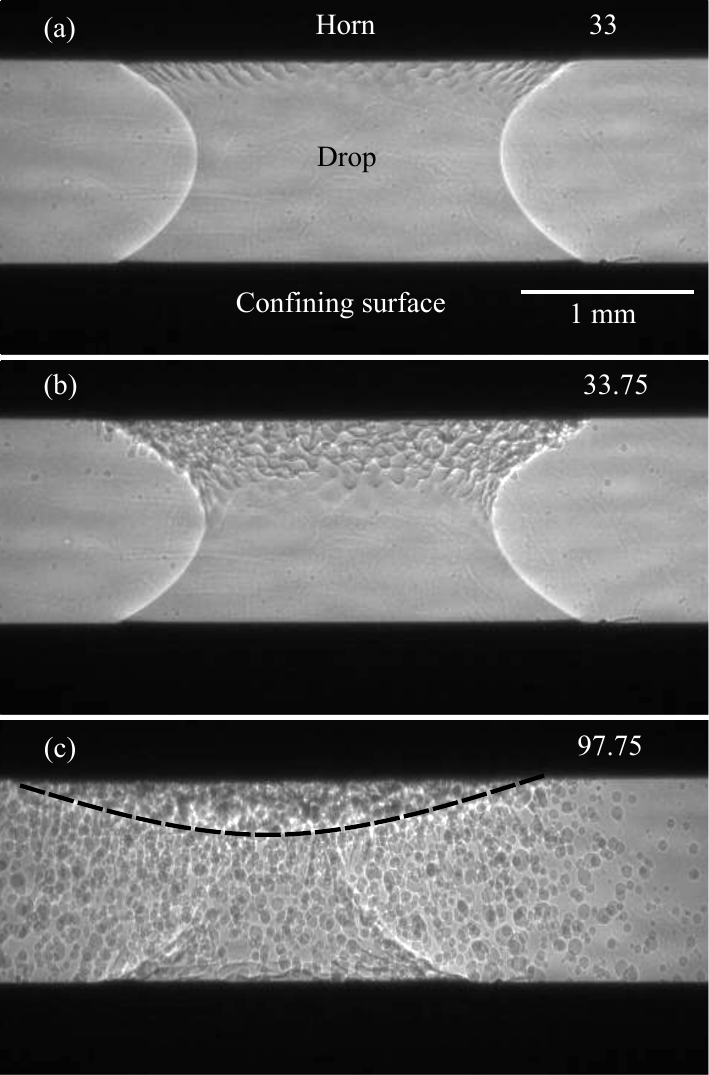}
    \caption{The dynamics of a confined water drop with $R_c = 1.4 \; \si{\milli\meter}$ and $h = 1.3 \; \si{\milli\meter}$. The snapshots of the X-ray phase-contrast images are labeled with their non-dimensional times, $t/T_d$. (a) The first waves generated in this drop correspond to subharmonic, azimuthal waves. (b) The subharmonic waves continue to grow and cover only the top half of the drop before the ejection of daughter droplets sets in. (c) Over time, only the top half of the confined drop ejects daughter droplets. The top half shows the same ejection dynamics as a free drop indicated by the black dashed line.}
    \label{fig: confined_drop}
\end{figure*}

\begin{figure*}[t]
    \centering
    \includegraphics[width=\textwidth]{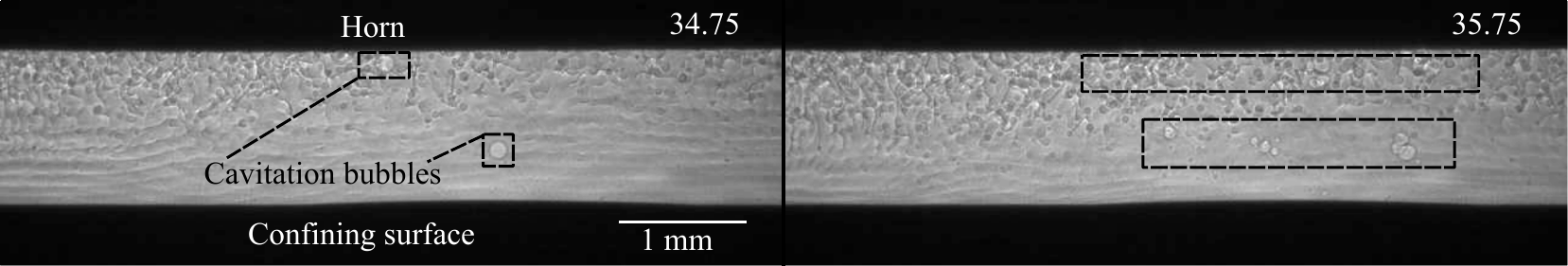}
    \caption{The dynamics of a confined drop with $R_c = 4.2 \; \si{\milli\meter}$ and $h = 1.3\;\si{\milli\meter}$. The snapshots of the X-ray phase-contrast images are labeled with their non-dimensional times, $t/T_d$. The image on the left depicts one of the first instants when cavitation bubbles become visible, enclosed in black boxes here. After 50 \si{\micro\second} (1 period), more cavitation bubbles have been nucleated, which are also enclosed in black boxes on the right image.}
    \label{fig: confined_drop_cavitation}
\end{figure*}

\begin{figure*}[t]
    \centering
    \includegraphics[width=0.8\textwidth]{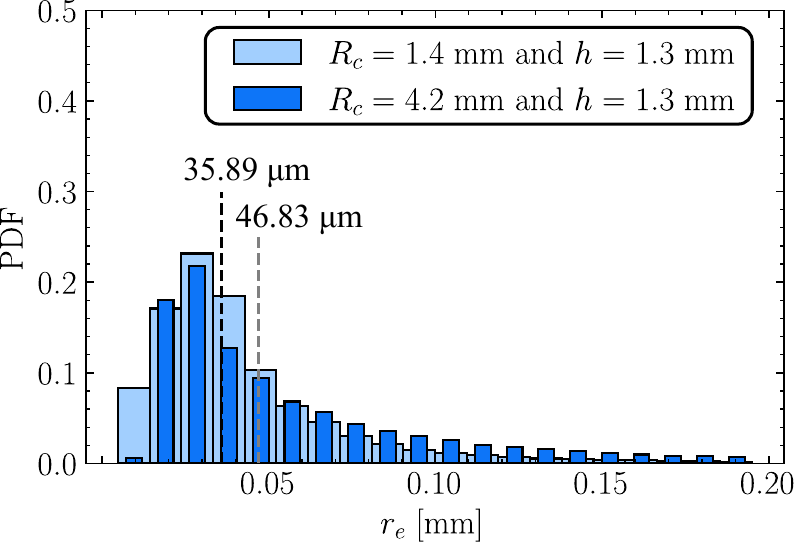}
    \caption{The probability distribution function of the radius of the ejected droplets plotted for the two confined drop cases. For $R_c = 1.4\;\si{\milli\meter}$, the PDF is plotted for $65 \leq t/T_d \leq 185$, while for $R_c = 4.2\;\si{\milli\meter}$, the PDF is plotted for $25 \leq t/T_d \leq 150$. The mean ejected droplet sizes for both cases are indicated with dashed lines (black for $R_c = 1.4\;\si{\milli\meter}$ and gray for $R_c = 4.2\;\si{\milli\meter}$).}
    \label{fig: confined_drop_histogram}
\end{figure*}

A confined, wetting water drop is depicted in Fig.~\ref{fig: confined_drop}. It has a contact radius of $R_c \approx 1.4 \; \si{\milli\meter}$, similar in volume to the free, pendant drop shown in Fig.~\ref{fig: wave_modes}. The distance between the active face of the horn and the rigid surface is $h = 1.3\;\si{\milli\meter}$. Using the convention provided by \cite{Moussatov2005}, this corresponds to $h/\lambda_f = 0.018$ and $R_c/\lambda_f = 0.017$, where $\lambda_f$ is the acoustic wavelength in water for $f_d = 20 \;\si{\kilo\hertz}$. For the above combination of $h/\lambda_f$ and $R_c/\lambda_f$ values, not enough pressure amplification is achieved in the liquid volume and no cavitation is generated. Instead, subharmonic, azimuthal waves are triggered without the appearance of axisymmetrical, harmonic waves. This is different from the pendant drop, where harmonic waves are triggered first, and subharmonic waves are generated later.

Interestingly, these waves cover only the top half of the drop, before they start ejecting daughter droplets. This could be simply due to the unstable nature of the wetting form of the confined water drop. The bridge breaks up and the lower half of the drop is disconnected from the top half. Fig.~\ref{fig: confined_drop}(c) shows that the ejection behavior of this drop is similar to a free pendant drop having characteristics equivalent to a drop shape indicated by the black dotted line. As the behavior is similar, it is to be expected that the ejected droplet size distribution for this case is equivalent to that of the pendant water drop (see Fig.~\ref{fig: confined_drop_histogram}).


Repeating the experiment with a larger drop shows some differences. Fig.~\ref{fig: confined_drop_cavitation} shows a confined drop with $R_c = 4.2 \; \si{\milli\meter}$. The interface for this configuration is out of the field of view of the image. The distance between the horn and the rigid surface is again maintained at $h = 1.3\;\si{\milli\meter}$. The behavior of the generated waves is the same as the free pendant drop with a large contact radius as described in Section~\ref{sec:water_drop}. Harmonic standing waves are coupled with subharmonic, semi-circular waves from point source locations in the horn. However, for this combination of $h/\lambda_f = 0.018$ and $R_c/\lambda_f = 0.056$, a sufficient pressure amplification is achieved within the liquid volume, to nucleate cavitation bubbles, as seen in Fig.~\ref{fig: confined_drop_cavitation}. These vapor bubbles oscillate and undergo multiple growth and collapse cycles along with the acoustic excitation and are found to affect the ejection process.      

The ejected droplet size distribution for both of the confined water drops - with and without cavitation - is plotted in Fig.~\ref{fig: confined_drop_histogram}. For $R_c = 4.2\;\si{\milli\meter}$, the ejected droplet size distribution was estimated manually due to the difficulty of distinguishing the waves from the ejected droplets by the edge detection algorithm.  As can be seen from the figure, the overall trend of the distribution is similar for both cases. This is unsurprising given that the ejection still occurs as a result of the breakup of capillary waves, irrespective of whether cavitation occurs within the liquid or not. However, there is a slight increase in the number of larger ejected droplets ($r_e > 0.05\;\si{\milli\meter}$) for the case when cavitation occurs ($R_c = 4.2\;\si{\milli\meter}$), shifting the mean of the distribution to a higher value. This suggests that a larger size range exists for droplets produced by the combined effect of cavitation and capillary waves as compared to droplets ejected purely by capillary waves, confirming the hypothesis of \cite{antonevich1959}. A possible explanation for the variation could be the high accelerations associated with the oscillation of the cavitation bubbles within the liquid leading to the formation of higher amplitude capillary waves, which produce droplets of different sizes when they break up, as compared to the capillary waves that are unaffected by the presence of the cavitation bubbles. 

A clear distinction cannot be made between the effect of the capillary waves, the drop confinement, and the cavitation-related activities on the ejected drop sizes. However, it must be noted that the only notable difference between $R_c = 1.4\;\si{\milli\meter}$ and $R_c = 4.2\;\si{\milli\meter}$ is the creation of vapor bubbles due to cavitation, and this is reflected in the ejected drop size distribution. While the case of $R_c = 1.4\;\si{\milli\meter}$ is directly comparable to a free pendant water drop, the case of $R_c = 4.2\;\si{\milli\meter}$ has a similar trend but with marked differences to the standard pendant drop case. 

\subsubsection{Bubble entrapped within a drop}
\label{sec:confined_bubble_drop}

\begin{figure*}[p]
    \centering
    \includegraphics[width=\textwidth]{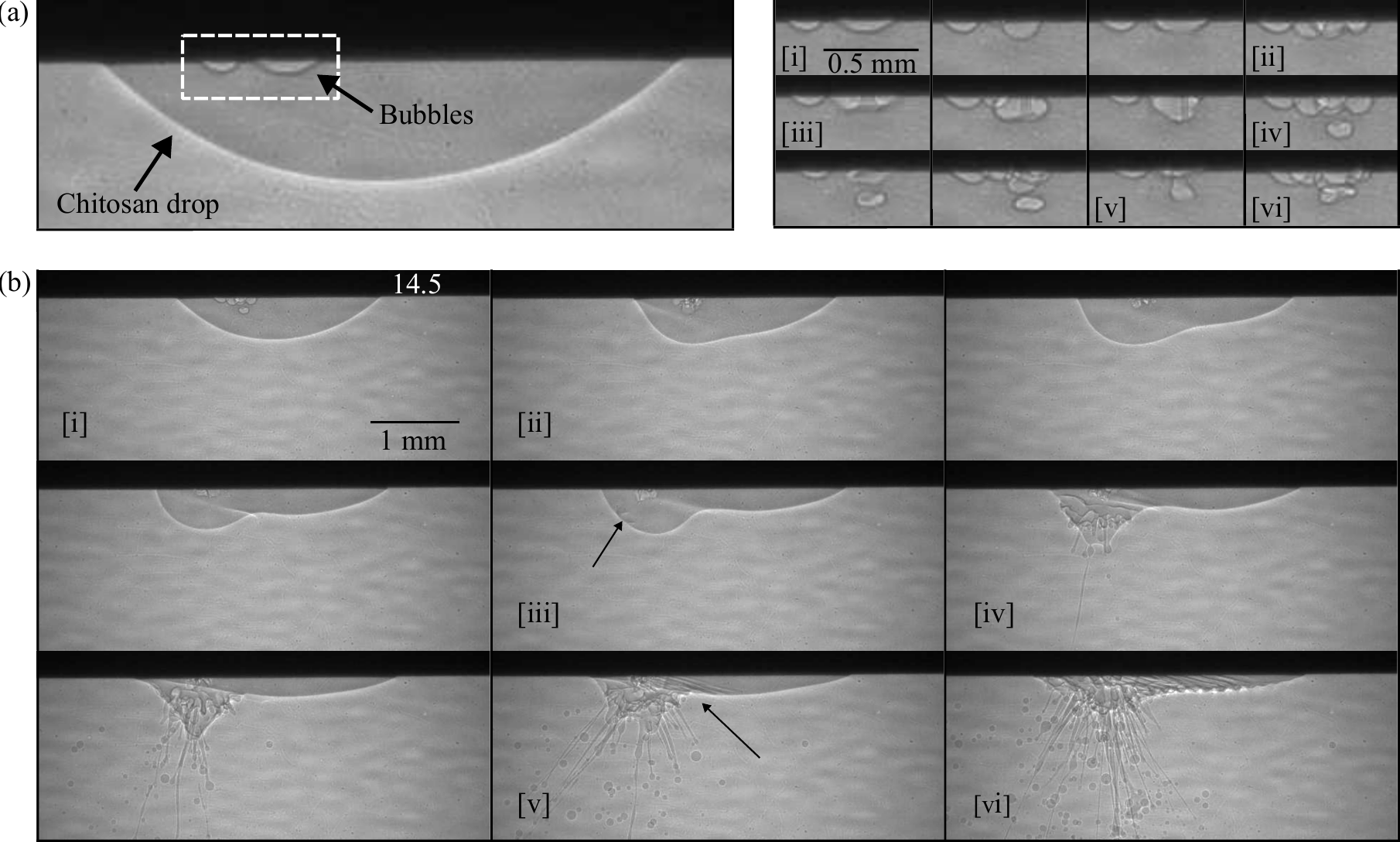}
    \caption{(a) The left frame shows two bubbles entrapped within a chitosan drop. The frames on the right show the zoomed-in section of the square box on the left frame. The inter-frame time is $t/T_d = 0.5$ with frame [i] corresponding to $t = 0$. [ii] The violent collapse of one of the bubbles leads to the formation of a jet and other bubbles. [iii] The jet is clearly visible here [iv] Continuous jetting leads to the formation of daughter bubbles. [v] The daughter bubble oscillates in tandem with the attached bubble [vi] the interaction between these bubbles causes further jets (also seen in [v]), and multiple daughter bubbles are created. (b) The evolution of the drop interface in tandem with the trapped bubble activity. The interframe time is $t/T_d = 25$. Frame [i] shows the system at $t/T_d = 14.5$. [ii] The drop bulges to the left and migrates as the bubbles translate along the horn face [iii] Initial ripples on the drop surface (black arrow), as it experiences Rayleigh-Taylor instability due to high radial accelerations [iv] Higher modes of the spherical Rayleigh-Taylor instability on the drop surface [v] Ejection of ligaments and droplets with the development of subharmonic waves in the part of the drop without the bubbles (black arrow) [vi] Co-evolution of ligament ejection on the left and subharmonic waves on the right of the pendant drop}
    \label{fig: bubble_behavior_in_drop}
\end{figure*}

The second configuration used to study the effect of cavitation on the ejection process involves trapping air bubbles inside a droplet. Air and 2\% chitosan are injected together using the microsyringe and placed on the active face of the ultrasonic horn, inducing trapped bubbles inside the drop. The qualitative results discussed here can equally be translated to bubbles trapped in water drops, apart from the viscoelastic effects. However, maintaining such an unstable configuration with a bubble-in-water drop, long enough to perform the X-ray imaging was impossible, and therefore, the more stable configuration employing chitosan drops was used instead. The initial configuration obtained is as shown in Fig.~\ref{fig: bubble_behavior_in_drop}(a), with the frames on the right showing the behavior of the trapped bubbles upon acoustic excitation. We will discuss the behavior of the larger bubble shown in Fig.~\ref{fig: bubble_behavior_in_drop}(a) as it has a radius close to the resonant size, $R \approx R_\mathrm{res} = 150 \; \si{\micro\meter}$, where $R_\mathrm{res}$ is the resonant bubble size for $f_d = 20 \; \si{\kilo\hertz}$ as predicted from linearizing the Rayleigh-Plesset equation \citep{Brennen2013}. 

Initially, this bubble oscillates in the vertical direction along with the applied excitation. After a few oscillation cycles, the bubble jets toward the active face of the horn as seen in Fig.~\ref{fig: bubble_behavior_in_drop}(a)[ii] and [iii]. A higher shape mode instability is developed (purely zonal, corresponding to $(k,l) = (4,0)$ according to the terminology used in \cite{Ding2022}) before a daughter bubble is pinched off (seen in Fig.~\ref{fig: bubble_behavior_in_drop}[iv]). The daughter bubble oscillates in tandem with the primary bubble and the out-of-phase oscillation of this bubble creates a highly focused jet toward the primary bubble. Over time, multiple bubbles are pinched off as the surface tension of the bubble is unable to withstand the increasing acceleration of the horn. These bubbles oscillate together, contributing to the interface dynamics of the drop. 

The radial acceleration created by the trapped bubble oscillations leads to the formation of a bulge in the drop interface on the left side as seen in Fig.~\ref{fig: bubble_behavior_in_drop}(b)[ii]. This bulge translates along the drop interface as the bubbles translate along the surface of the horn. The high acceleration appears to break the drop into two sections, with the part of the drop with the bubbles and the bulge showing a different behavior as compared to the rest of the drop. The continued cycles of growth and collapse of the bubbles create large radial accelerations within the liquid volume, leading to the onset of the Rayleigh-Taylor instability on the bulged part of the droplet, with the initial ripples indicated by the black arrow in Fig.~\ref{fig: bubble_behavior_in_drop}(b)[iii]. The geometry of the problem implies that this can be classified as a spherical Rayleigh-Taylor instability \citep{Plesset1954}. It must also be mentioned that the high viscoelasticity of the chitosan solution stabilizes the drop interface against the Rayleigh-Taylor instability \citep{Prosperetti1977, Zeng2018}, and only after a certain threshold amplitude is exceeded does the fluid yield and the drop interface becomes unstable. Due to the amplitude of the horn increasing over time during the transient period, the Weber ($\rho u_\mathrm{in}^2 R_c/\sigma$, where $u_\mathrm{in}$ is the time-dependent velocity of the drop interface) and Reynolds numbers ($\rho u_\mathrm{in} R_c/\mu_\mathrm{eff}$) at the interface increase with time, causing the initial ripples to appear. With further increase in the amplitude over time, higher modes are set in, creating more ripples on the drop interface (see Fig.~\ref{fig: bubble_behavior_in_drop}(b)[iv]).  

The instability of the interface leads to the ejection of ligaments and daughter droplets in this part of the pendant drop. The ejection process here starts at much earlier driving cycles as compared to a pendant drop without cavitation (refer to Fig.~\ref{fig: chitosan_waves_ejection_entrainment}). This implies that the large inertia and accelerations created by cavitation-related activities hasten the onset of ejection in the pendant drop. These accelerations are also responsible for the direct ejection of daughter droplets along with the ligaments. Furthermore, the ligaments eject satellite droplets due to the Rayleigh-Plateau instability. Both of these scenarios are not present when there are no trapped bubbles in the chitosan drop as seen in Fig.~\ref{fig: chitosan_waves_ejection_entrainment}. The consequences of having trapped bubbles are discussed further in Section~\ref{sec:disc}. 

The part of the drop without the bubble is still attached to the horn and shows the classic behavior as described previously for the other pendant drops. The black arrow in Fig.~\ref{fig: bubble_behavior_in_drop}(b)[v] shows the onset of subharmonic capillary waves, which travel and spread across the whole attached drop as seen in Fig.~\ref{fig: bubble_behavior_in_drop}(b)[vi]. The waves appear to be inclined because of the tilted orientation of the drop on the face of the horn. Since the whole drop is in focus in the phase-contrast images, the projection of a tilted drop onto a plane while imaging makes it look like the waves themselves are inclined, even though this is not the case. The waves in this part of the drop co-evolve with the ejecting part and transition to ejecting ligaments at similar driving cycles as the chitosan drop without any trapped bubbles (see Section~\ref{sec:chitosan}).

\section{Discussion}
\label{sec:disc}

A variety of characteristics are found in the drop atomization process based on the different tested configurations. The dependence on the fluid-structure interaction was visualized clearly when a drop with a larger contact radius with the horn was used. While fluid-structure interaction in this context was briefly studied by \cite{James2003}, the fundamental mode shape of their transducer implied that they could not see localized variations on the drop surface. With the fundamental mode of a nonlinear ultrasonic horn, however, it is visibly clear that the ejection onset is affected. While this did not have an effect on the sizes of the ejected droplets in our case, further study is required to characterize whether localized regions of high-displacement on the surface of the transducer also have an impact on the products of atomization. The interaction could then be used in a beneficial manner to create daughter droplets with lower input power. 

The ejected droplet size distribution was estimated for water drops. While the overall distribution is polydisperse, the median size of ejected droplets correlated well with theoretical predictions. This implies that a very quick estimate of the ejected droplet size can be calculated from the driving frequency and the fluid properties. The transition from ejecting smaller to larger droplets over time implies that nonlinear effects dominate after a certain point. Further research is required to distinguish when and how this transition occurs.

Water drops were confined between the ultrasonic horn and the rigid surface to utilize the ability of the horn to generate cavitation in confined thin liquid layers. Only certain combinations of $h/\lambda_f$ and $R_c/\lambda_f$ can produce the necessary pressure amplification to nucleate vapor bubbles within the liquid. No cavitation was produced for the case of $R_c = 1.4\;\si{\milli\meter}$, but the confinement alone affected the atomization. The configuration generated subharmonic capillary waves only within the top half of the drop and only this part ejected daughter droplets. It would be curious to see if similar capillary wave formation and ejection behavior is noticed for drops that are non-wetting as they would fundamentally have a different shape to the drop depicted in Fig.~\ref{fig: confined_drop}. For $R_c = 4.2\;\si{\milli\meter}$ on the other hand, the pressure amplification within the droplet was large enough to cavitate vapor bubbles that experience multiple growth and collapse cycles and create large amplitude capillary waves. As mentioned previously, it is difficult to distinguish between the effects of confinement and cavitation in the ejection process. However, our results suggest that the confinement alone is insufficient to change the size distribution of the ejected droplets while cavitation seems to slightly modify the daughter droplet size distribution. Optimizing the $h/\lambda_f$ and $R_c/\lambda_f$ could create localized regions of high-pressure \citep{Moussatov2005}, which provides another method to modify the size distribution of the daughter droplets.   

Using a viscoelastic drop to study the atomization process showed interesting features. The requirement of higher input amplitude to atomize viscoelastic drops was shown to correspond quite well with the existing theory. Extended ligament formation and air entrainment within the primary drop were visualized succinctly. Again, the oscillation of the trapped bubbles due to entrainment has an effect on the ejection process. From the Supplementary Video corresponding to Fig.~\ref{fig: chitosan_waves_ejection_entrainment}, it seems that the ejection of ligaments and daughter droplets is quickened by the oscillation of the entrained bubbles in tune with the acoustic excitation. The case of the initially trapped air bubbles within the chitosan drop provides further proof that this is true. The onset of ejection here is faster as compared to a pure chitosan drop due to the presence of trapped bubbles that generate high radial accelerations and accelerate the drop interface. This implies that cavitation and cavitation-related activities can easily lower the threshold required to eject daughter droplets, even though the generation of cavitation within the liquid, and the resulting products of ejection are not controllable. For both chitosan cases, it can be seen qualitatively that a wide range of droplet sizes are ejected. Furthermore, quite a few drop-encapsulated bubbles that are stable for relatively long periods of time are ejected as well, which raises the question of the homogeneity of the atomization products. This could have consequences in several applications involving atomization such as inkjet printing, fertilizer sprays, or pulmonary drug delivery. 

The configurations used here are quite limited to elucidate all the parameters in play for the atomization process. It would be interesting to see how the dynamics change as the input power and driving frequency are changed as well. While some differences between capillary and cavitation-based ejection have been highlighted, the current setup does not fully allow us to separate the effects and comment on the major distinctions brought about by the two different phenomena. Further research needs to be carried out to study the performance of only capillary or cavitation-based ejection, and the conjunction of both of them. For viscoelastic drops, it would be interesting to separate the role of the viscous and the elastic effects in the atomization process, especially when air entrainment and cavitation-related activities are involved. The interplay between the viscous, capillary, and elastic effects in stabilizing the primary drop interface and the ligaments could play a crucial role in determining the nature of the ejection products and further open up a wide range of applications. If the dynamics are well understood, it could be beneficial to dilute or concentrate liquid solutions with polymers to obtain the exact type of ejection behavior required for a certain application, which makes it a promising avenue to explore. 

Finally, we would like to highlight the benefits of performing spatiotemporally resolved X-ray phase-contrast imaging for these experiments. The advanced imaging technique proved critical for most of our reported findings. It allowed us to show that cavitation does not occur inside a free pendant drop and helped us highlight the consequences of cavitation and large bubble oscillations on the drop atomization process. 

\section{Conclusion}
\label{sec:conc}

In this study, we characterize the atomization behavior of pendant drops subjected to ultrasonic excitation using advanced experimental techniques. The goal was to study the dependence of drop atomization on fluid properties, such as drop volume and viscosity, by performing spatiotemporally resolved imaging measurements and leveraging the virtual infinite depth of field of X-ray phase-contrast imaging as compared to conventional imaging. Our results also shed light on the modifications to the atomization of the drop, subject to the generation of cavitation and cavitation-related activities. Experiments were carried out with water and chitosan drops. Atomization was found to be dictated by the creation of surface waves on the drop. Both harmonic and subharmonic waves were observed with the time needed for the subharmonic waves to set in the drop being greater than the time needed to generate the harmonic waves. The observed transition from one wave mode to another was dictated by the fluid-structure interaction between the pendant drop and the horn. This was evident when drops had a large contact radius (water drop with $R_c = 2.8\;\si{\milli\meter}$), as well as when they had a pronounced dynamic contact line and subsequently underwent spreading (chitosan drop with $R_c = 1.6\;\si{\milli\meter}$). The larger pendant water drop experienced a quicker onset of ejection due to this fluid-structure interaction.

The amplitude of the surface waves progressively increases over time, creating crests and troughs on the surface of the primary drop. Once the identified threshold displacement is reached, the oscillating pendant drop atomizes and ejects daughter droplets. The threshold value was determined from scaling relations that were different based on the inviscid or highly viscous nature of the fluid. The median and mean ejected droplet sizes for the pendant water drop correspond well with the predictions from previous experiments as well as theory \citep{Lang1962}.

In addition, configurations were devised to study the effect of cavitation and the combined effects of both cavitation and capillary waves on drop atomization. Drops were confined between the horn and a rigid surface or bubbles were initially trapped within the drop volume. Although no apparent change in the mean daughter droplet size was noticed, the confined drop experiments indicate a much wider range of ejected droplets provided cavitation is generated within the drop. This could be attributed to the formation of high-amplitude capillary waves from the radial oscillations of the generated vapor bubbles that slightly increase the sizes and range of the ejected droplets. However, inducing cavitation within the drop cannot be controlled or localized in this case.

Finally, when entrapped air bubbles were introduced in the chitosan drop, the cavitation-related activities were observed to substantially quicken the onset of ejection in contrast to having only surface waves. The large radial accelerations generated by the expansion and collapse of the bubbles resulted in an unstable drop interface, which led to the formation of the Rayleigh-Taylor instability. The instability stimulated an earlier ejection of ligaments and daughter droplets from the pendant drop. Both the cavitation-based drop atomization studies imply that with cavitation, the creation of daughter droplets is easier. However, controlling the ejected size distribution and the end products with cavitation proved challenging as compared to having only capillary waves on the surface of the drop. Although further research is required if one wants to use cavitation to accelerate the ejection process and obtain controlled products from it, this work establishes promising avenues based on rich results.

\section*{CRediT authorship contribution statement}

\textbf{Anunay Prasanna} - Methodology, Formal analysis, Investigation, Data curation, Visualization, Writing - original draft, Writing - review \& editing;
\textbf{Luc Biasiori-Poulanges} - Conceptualization, Methodology, Investigation, Writing - review \& editing;
\textbf{Ya-Chi Yu} - Formal analysis, Investigation, Writing - review \& editing;
\textbf{Hazem El-Rabii} - Conceptualization, Writing - review \& editing;
\textbf{Bratislav Luki\'c} - Methodology, Resources, Writing - original draft, Writing - review \& editing;
\textbf{Outi Supponen} - Conceptualization, Methodology, Resources, Investigation, Visualization, Writing - review \& editing, Supervision, Project administration, Funding acquisition

\section*{Declaration of competing interest}
The authors declare that they have no known competing financial or personal interests that could have influenced the work reported in this manuscript.

\section*{Acknowledgements}

The authors would like to acknowledge the Swiss National Science Foundation (SNSF project grant number 200021\_200567), ETH Z{\"u}rich Postdoctoral Fellowship, ETH Z{\"u}rich, and the European Synchrotron Radiation Facility. The results presented here were gathered during the allocated proposal beamtime ME-1599 on beamline ID19. The authors would like to thank Dr.\ Claire Bourquard for helping with the preparation of the chitosan solution and Dr.\ Dhananjay Deshmukh for his help with the laser scanning vibrometry experiments. The gratitude is also extended to colleagues at McGill University, Dr.\ Zhenwei Ma and Prof.\ Jianyu Li, for giving the initial idea to the authors thanks to their accidental observations of drop atomization.

\appendix

\section{Preparation of 2\% chitosan solution}
\label{app:chitosan}

To make the 2\% chitosan solution (10 mL), 200 \si{\milli\gram} of low molecular weight chitosan powder (Sigma Aldrich) is added to 9 \si{\milli\liter} of deionized water. The solution is stirred to obtain a suspension of chitosan particles. Then, 1 \si{\milli\liter} of 1M HCl solution (Sigma Aldrich) is added into the suspension. The chitosan-HCl-water solution is then left overnight on a rotating mixer to allow the complete dissolution of chitosan and to obtain a homogeneous solution.

\section{Droplet size distribution for pendant water drop with $R_c = 2.8\;\si{\milli\meter}$}
\label{app:dropSize}

\begin{figure*}[t]
    \centering
    \includegraphics[width=0.7\textwidth]{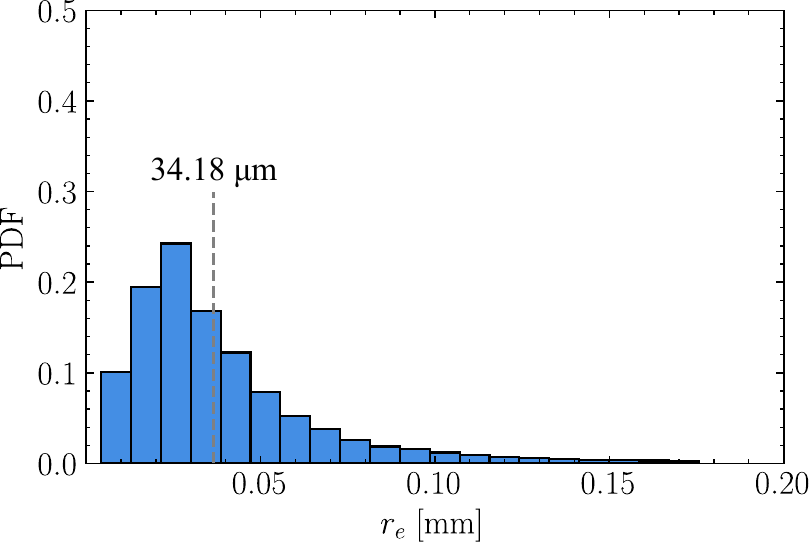}
    \caption{The probability distribution function for the ejected droplets of the pendant water drop with $R_c = 2.8 \; \si{\milli\metre}$ plotted for driving cycles, $70 \leq t/T_d \leq 125$. The mean value (indicated by the dashed line) is 34.18 $\pm$ 9.52 \si{\micro\meter} and the median value is 29.6 $\pm$ 9.52 \si{\micro \meter}}
    \label{fig: drop_size_2}
\end{figure*}

Fig.~\ref{fig: drop_size_2} plots the probability distribution function for the daughter droplets of the pendant water drop with $R_c = 2.8\;\si{\milli\meter}$. The distribution, median, and mean values are similar to those obtained for the smaller pendant drop. The median value predicted by Eq.~(\ref{eq:lambda}) is dependent only on the driving frequency and the fluid properties, and therefore, the similarity in the distribution is as expected. 


\bibliographystyle{elsarticle-harv} 
\bibliography{references}





\end{document}